%% file: main.tex
\documentclass[runningheads,envcountsame]{llncs}
\usepackage[T1]{fontenc}
\usepackage{amssymb}
\usepackage{amsmath}
\usepackage{amsfonts}
\usepackage{mathabx}
\usepackage[ligature]{semantic}
\usepackage{mathtools}
\usepackage{latexsym}
\usepackage{stmaryrd}
\usepackage{cite}
\usepackage{xspace}
\usepackage{graphicx}
\DeclareGraphicsExtensions{.pdf,.png,.jpg}
\usepackage{tikz}
\usetikzlibrary{shapes,trees,arrows,positioning,calc}
\usepackage[scaled=.92]{helvet}
\usepackage{times}
\usepackage{listings}
\usepackage{enumerate}
\usepackage{caption}
\usepackage{subfigure}
\usepackage{url}

\usepackage[ruled,lined,longend]{algorithm2e}
\usepackage{tabularx}

\usepackage[amsfonts]{logic-thm}

\input{defs}
\begin{document}

\title{Learning Invariants using Decision Trees}

\author{Siddharth Krishna \and Christian Puhrsch \and Thomas Wies}

\institute{New York University, USA}

\maketitle

\input{abstract}
\input{introduction}

\input{example}
\input{preliminaries}
\input{algorithm}
\input{results}
\input{related}

%\input{conclusion}

\bibliographystyle{abbrv}
\bibliography{main}

\end{document}

%% file: defs.tex
\newcommand{\smartparagraph}[1]{\smallskip\noindent
{\bf #1}\ }

\renewcommand{\m}[1]{\mathsf{#1}}

\newcommand{\pset}[2]{\set{\,#1\mid#2\,}}

\newcommand{\lfp}{\m{lfp}}

\newcommand{\init}{\m{init}}
\newcommand{\safe}{\m{safe}}
\newcommand{\error}{\m{error}}
\newcommand{\post}{\m{post}}
\newcommand{\pre}{\m{pre}}

\newcommand{\good}{\m{good}}
\newcommand{\bad}{\m{bad}}
\newcommand{\inv}{\m{inv}}

\newcommand{\jbasicstyle}{\sffamily}
\lstdefinelanguage{SPL}{
  morekeywords={acc,struct,if,else,returns,procedure,requires,
    ensures,:=,var,new,old,free,implicit,invariant,modifies,
    call,locals,assume,assert,choose,havoc,ghost,predicate,
    function,return,while},
  deletekeywords={union,int},
  numbers=left,
  numberstyle=\tiny,
  basicstyle=\jbasicstyle,
  columns=flexible,
  morecomment=*[s][\textsl]{/*:}{*/},
  morecomment=*[l][\textsl]{//:},
  mathescape=true,
}

%% file: abstract.tex
\begin{abstract}
  The problem of inferring an inductive invariant for verifying
  program safety can be formulated in terms of \emph{binary
    classification}. This is a standard problem in machine learning:
  given a sample of good and bad points, one is asked to find a
  classifier that generalizes from the sample and separates the two
  sets. Here, the good points are the reachable states of the program,
  and the bad points are those that reach a safety property violation. Thus, a
  learned classifier is a candidate invariant. In this paper, we
  propose a new algorithm that uses decision trees to learn candidate
  invariants in the form of arbitrary Boolean combinations of
  numerical inequalities. We have used our algorithm to verify C
  programs taken from the literature. The algorithm is able to infer
  safe invariants for a range of challenging benchmarks and compares
  favorably to other ML-based invariant inference techniques. In
  particular, it scales well to large sample sets.

  %We also plan to extend this approach to non-linear and hybrid systems.
\end{abstract}
%%% Local Variables:
%%% mode: latex
%%% TeX-master: "main"
%%% End:

%% file: introduction.tex
\section{Introduction}

Finding inductive invariants is a fundamental problem in program
verification. Many static analysis techniques have been proposed to
infer invariants automatically. However, it is often difficult to
scale those techniques to large programs without compromising on
precision at the risk of introducing false alarms. Some techniques,
such as abstract
interpretation~\cite{CousotCousot77AbstractInterpretation}, are
effective at striking a good balance between scalability and precision
by allowing the analysis to be fine-tuned for a specific class of
programs and properties. This fine-tuning requires careful engineering
of the analysis~\cite{DBLP:conf/esop/CousotCFMMMR05}. Instead
of manually adapting the analysis to work well across many similar
programs, refinement-based techniques adapt the analysis automatically
to the given program and property at hand~\cite{DBLP:conf/cav/ClarkeGJLV00}.
A promising approach to achieve this automatic adaptation is to
exploit synergies between static analysis and
testing~\cite{DBLP:conf/sigsoft/GulavaniHKNR06,
  DBLP:conf/issta/YorshBS06,
  DBLP:conf/popl/GodefroidNRT10}. Particularly interesting is the use
of Machine Learning (ML) to infer likely invariants from test
data~\cite{SNA12,SGHAN13,DBLP:conf/cav/0001LMN14}. In this paper, we
present a new algorithm of this type that learns arbitrary Boolean
combinations of numerical inequalities.

In most ML problems, one is given a small number of sample points
labeled by an unknown function. The task is then to learn a classifier
that performs well on unseen points, and is thus a good approximation
to the underlying function. Binary classification is a specific
instance of this problem. Here, the sample data is partitioned into
good and bad points and the goal is to learn a predicate that
separates the two sets. Invariant inference can be viewed as a binary
classification problem~\cite{SGHAN13}. If the purpose of the invariant
is to prove a safety property, then the good points are the
forward-reachable safe states of the program and the bad points are
the backward-reachable unsafe states. These two sets are sampled
using program testing. The learned classifier then represents a
candidate invariant, which is proved safe using a static analysis or
theorem prover. If the classifier is not a safe invariant, the failed
proof yields a spurious counterexample trace and, in turn, new test
data to improve the classifier in a refinement loop.

Our new algorithm is an instance of this ML-based refinement scheme,
where candidate invariants are inferred using a decision tree learner.
In this context, a decision tree (DT) is a binary tree in which each
inner node is labeled by a function $f$ from points to reals, called a
\emph{feature}, and a real-valued \emph{threshold} $t$. Each leaf of
the tree is labeled with either ``good'' or ``bad''. Such a tree
encodes a predicate on points that takes the form of a Boolean
combination of inequalities, $f(\vec{x}) \leq t$, between features and
thresholds. Given sets of features and sample points,
a DT learner computes a DT that is consistent with the samples. In our
algorithm, we project the program states onto the numerical program
variables yielding points in a $d$-dimensional space. The features
describe distances from hyperplanes in this space. The DT learner thus infers
candidate invariants in the form of arbitrary finite unions of
polyhedra. However, the approach also easily generalizes to features
that describe nonlinear functions. Our theoretical contribution is a
probabilistic completeness guarantee. More precisely, using the
Probably Approximately Correct model for learning~\cite{Valiant84}, we
provide a bound on the sample size that ensures that our algorithm
successfully learns a safe inductive invariant with a given
probability.

We have implemented our algorithm for specific classes of features
that we automatically derive from the input program. In particular,
inspired by the octagon abstract
domain~\cite{DBLP:journals/lisp/Mine06}, we use as features the set of
all hyperplane slopes of the form $\pm x_i \pm x_j$, where $1 \leq i <
j \leq d$. We compared our implementation to other invariant
generation tools on benchmarks taken from the literature. Our
evaluation indicates that our approach works well for a range of
benchmarks that are challenging for other tools. Moreover, we observed
that DT learners often produce simpler invariants and scale better to
large sample sets compared to other ML-based invariant inference
techniques such
as~\cite{SNA12,SGHAN13,DBLP:conf/cav/0001LMN14,DBLP:conf/cav/0001A14}.

%% file: example.tex
\section{Overview}

\begin{figure}[t]

\begin{minipage}{.38\linewidth}
\begin{lstlisting}[language=SPL,escapeinside={@}{@},xleftmargin=2em]
var x, y: Int;
assume x @$=$@ 0 @$\land$@ y @$\neq$@ 0;

while (y @$\neq$@ 0) { @\label{code:loop}@
  if (y @$<$@ 0) {
    x := x - 1; y := y + 1;
  } else {
    x := x + 1; y := y - 1;
  }
}

assert x @$\neq$@ 0; @\label{code:assert}@
\end{lstlisting}
\end{minipage}%
\begin{minipage}{.62\linewidth}
  \includegraphics[width=\linewidth]{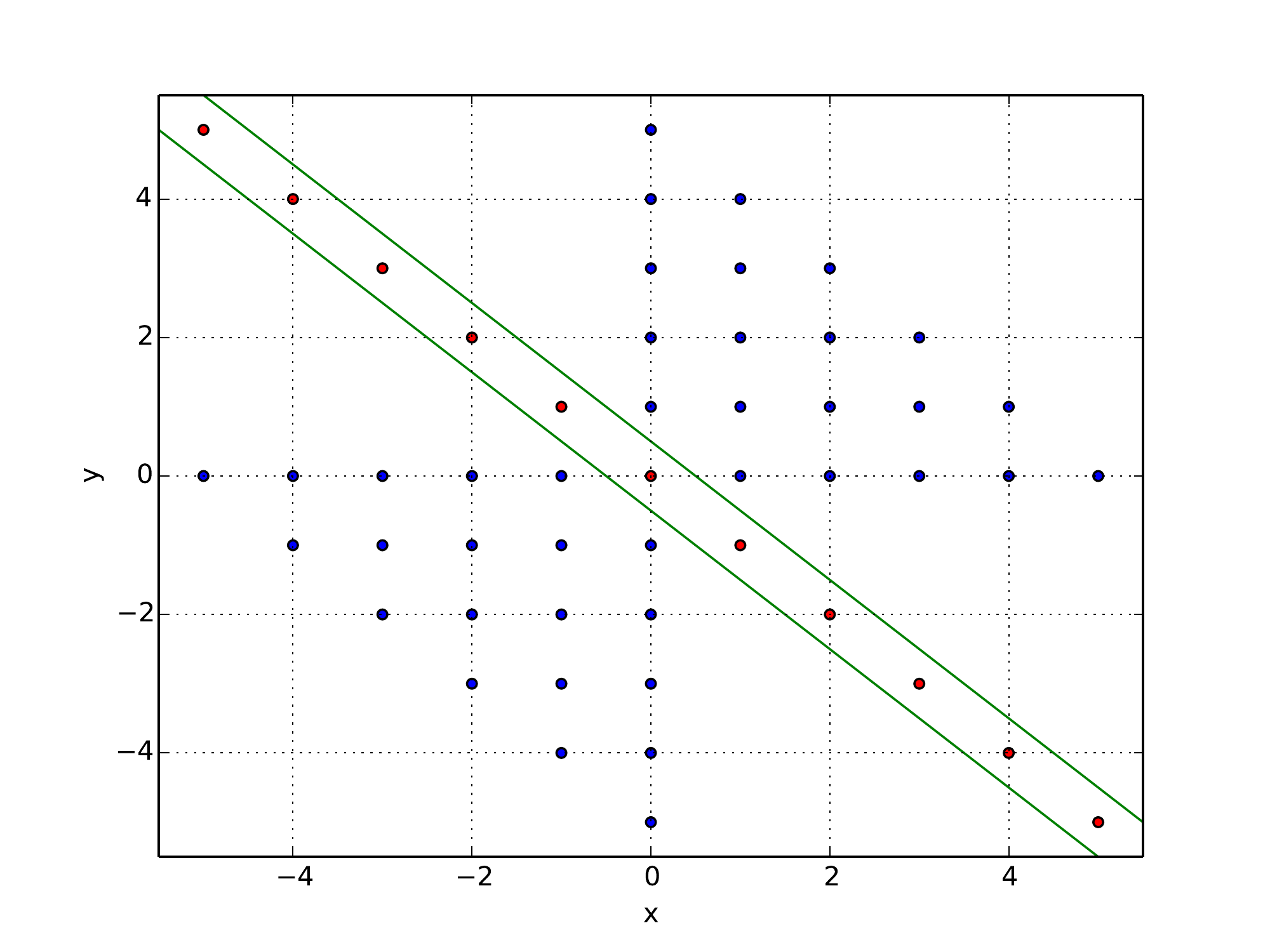}
\label{fig:example-good-bad}
\end{minipage}
\vspace*{-2em}
\caption{Example program. The right side shows some of the good and
  bad states of the program, in blue and red respectively\label{fig:program}}
\end{figure}

In this section, we discuss an illustrative example and walk through
the steps taken in our algorithm to compute invariants. To this end,
consider the program in Fig.~\ref{fig:program}. Our goal is to find an
inductive invariant for the loop on line~\ref{code:loop} that is
sufficiently strong to prove that the assertion $x \neq 0$ on
line~\ref{code:assert} is always satisfied.

\smartparagraph{Good and Bad States.}
We restrict ourselves to programs over integer variables $\vec{x} =
(x_1,\dots,x_d)$ without procedures. Then a \emph{state} is a point in
$\ZZ^d$ that corresponds to some assignment to each of the
variables. For simplicity, we assume that our example program has a
single control location corresponding to the head of the loop. That
is, its states are pairs $(v_1,v_2)$ where $v_1$ is the value of
\lstinline+x+ and $v_2$ the value of \lstinline+y+. When our program
begins execution, the initial state could be $(0, 1)$ or $(0, -3)$,
but it cannot be $(2, 3)$ or $(0, 0)$ because of the precondition
specified by the \lstinline+assume+ statement in the program. A
\emph{good state} is defined as any state that the program could
conceivably reach when it is started from a state consistent with the
precondition. If we start execution at $(0, -3)$, then the states we
reach are $\{(-1, -2), (-2, -1), (-3, 0)\}$ and thus these are all
good states.

Similarly, \emph{bad states} are defined to be the states such that if
the program execution was to be started at that point (if we ran the
program from the loop head, with those values) then the loop will exit
after a finite point and the assertion will fail. For example, $(0,
0)$ is a bad state, as the loop will not run, and we directly go to
the assertion and fail it. Similarly, $(-2, -2)$ is a bad state, as
after one iteration of the loop the state becomes $(-1, -1)$, and
after another iteration we reach $(0, 0)$, which fails the assertion.

The right-hand side of Fig.~\ref{fig:program} shows some of the good
and bad states of the program. A safe inductive invariant can be
expressed in terms of a disjunction of the indicated hyperplanes,
which separate the good from the bad states. Our algorithm
automatically finds such an invariant.

\smartparagraph{Overview.}
The high-level overview of our approach is as follows: good and bad
states are sampled by running the program on different initial
states. Next, a numeric abstract domain that is likely to contain the
invariant is chosen manually. We use disjunctions of octagons by
default. For each hyperplane (up to translation) in the domain we add
a new ``feature'' to each of the sample points corresponding to the
distance to that hyperplane. A Decision Tree (DT) learning algorithm
is then used to learn a DT that can separate the good and bad states
in the sample, and this tree is converted into a formula that is a
candidate invariant. Finally, this candidate invariant is passed to a
theorem prover that verifies the correctness of the invariant. We now
discuss these steps in detail.

%\todo{Mention that we could use the refinement loop here, and then in the
%implementation section say we didn't need to/didn't implement it?}

\smartparagraph{Sampling.}  The first step in our algorithm is to
sample good and bad states of this program. We sample the good states
by picking states satisfying the precondition, running the program
from these states and collecting all states reached. To sample bad
states, we look at all points close to good states, run the program
from these. If the loop exits within a bounded number of iterations
and fails the \lstinline+assert+, we mark all states reached as bad
states. The sampled good and bad states are shown in Fig.~\ref{fig:program}.  % We write them as rows in a matrix,
% \[\Xx =
% \begin{bmatrix}

% \end{bmatrix}
% \]
\iffalse
\begin{figure}
\includegraphics[width=\linewidth]{simple5.png}
\caption{Good and bad states, in blue and red respectively. The inequalities
used by the computed DT are also shown.}
\label{fig:example-good-bad}
\end{figure}
\fi

\smartparagraph{Features.}
The next step is to choose a candidate hyperplane set for the
inequalities in the invariant. For most of our benchmarks, we used the
\emph{octagon} abstract domain, which consists of all linear
inequalities of the form:
\[ b_i \cdot x_i + b_j \cdot x_j \leq c \text{ where } 1 \leq i < j
\leq d,\; b_i, b_j \in \{-1, 0,
1\}, \text{ and } c \in \ZZ.\] We then let $H =
\{\vec{w_1}, \vec{w_2}, \dotsc\}$ be the set of hyperplane slopes for this
domain. Then we transform our sample points (both good and bad) according to
these slopes. For each sample point $\vec{x}$, we get a new point
$\vec{z}$ given by $z_i = \vec{x} \cdot \vec{w_i}$. In our example,
the octagon slopes $H$ and some of our transformed good and bad points are
\begin{align*}
\small
H =
\begin{bmatrix*}[r]
  1 & 0 \\ 0 & 1 \\ 1 & -1 \\ 1 & 1 \\
\end{bmatrix*}, \qquad \text{and} \qquad
X \cdot H^T =
\begin{bmatrix*}[r]
0 & 1 \\
 1 & 0 \\
 -1 & 0 \\
\multicolumn{2}{c}{\dots} \\
0 & 0 \\
 2 & -2 \\
 -1 & 1 \\
\multicolumn{2}{c}{\dots} \\
\end{bmatrix*}
\cdot
\begin{bmatrix*}[r]
1 & 0 & 1 & 1 \\
0 & 1 & -1 & 1 \\
\end{bmatrix*}
=
\begin{bmatrix*}[r]
0 & 1 & -1 & 1 \\
 1 & 0 & 1 & 1 \\
 -1 & 0 & -1 & -1 \\
\multicolumn{4}{c}{\dots} \\
0 & 0 & 0 & 0 \\
 2 & -2 & 4 & 0 \\
 -1 & 1 & -2 & 0 \\
\multicolumn{4}{c}{\dots}
\end{bmatrix*} \enspace.
\end{align*}

\begin{figure}[t]
\centering
\includegraphics[width=.8\linewidth]{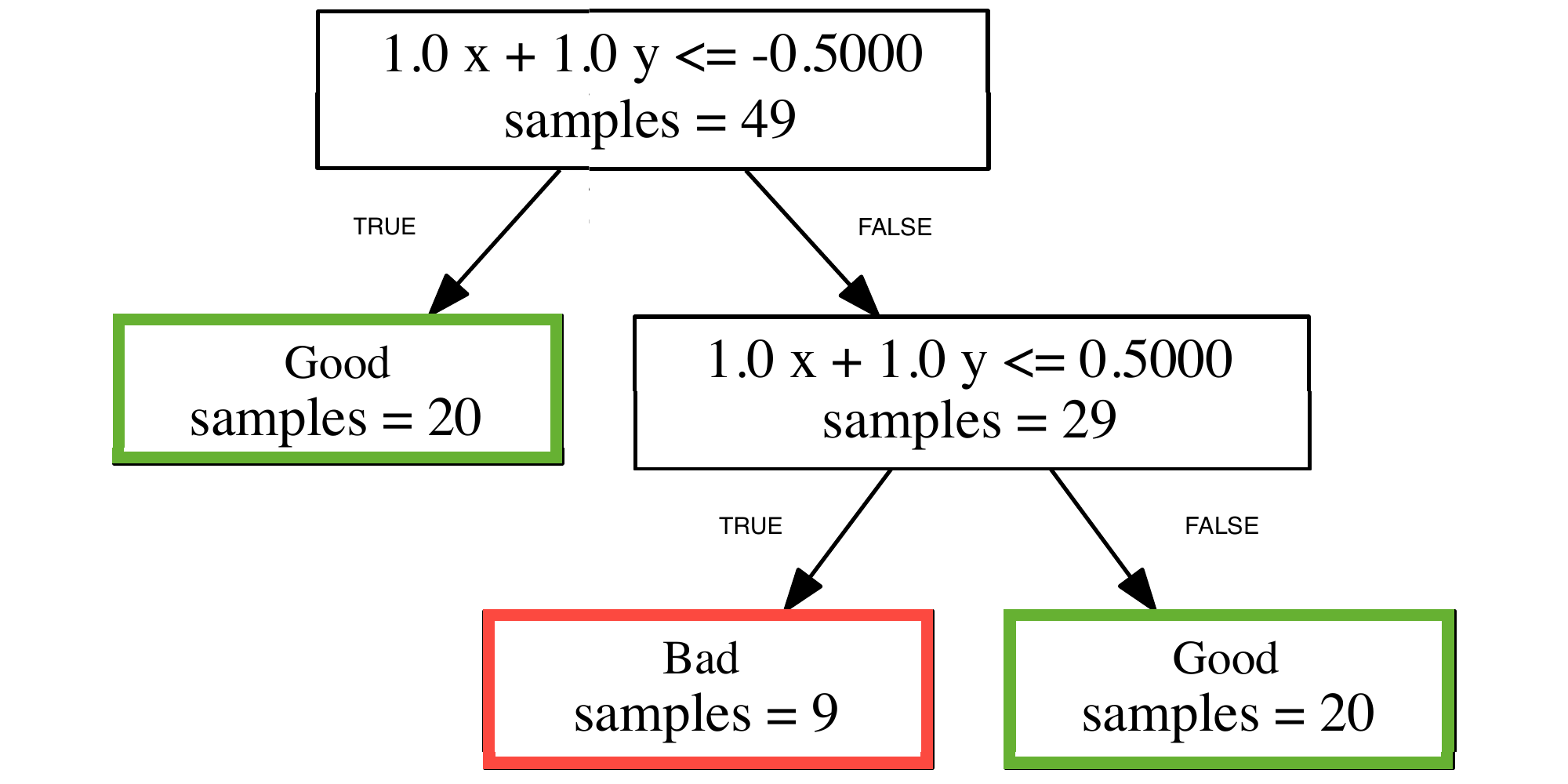}
\caption{Decision tree learned from the program and sample data in Fig.~\ref{fig:program}}
\label{fig:example-dtree}
\end{figure}

\smartparagraph{Learning the DT.}  After this transformation, we run a
Decision Tree learning algorithm on the processed data.  A DT (see
Fig.~\ref{fig:example-dtree} for an example) is a concise way to
represent a set of rules as a binary tree. Each inner node is labeled
by a feature and a threshold. Given a sample point and its features,
we evaluate the DT by starting at the root and taking the path given
by the rules: if the features is less than or equal to the threshold,
we go to the left child, otherwise the right. Leaves specify the
output label on that point. Most DT learning
algorithms start at an empty tree, and greedily pick the best feature
to split on at each node. From the good and bad states listed above,
we can easily see that a good feature to split on must be the last
one, as all bad states have the last column 0. Indeed, the first split
made by the DT is to split on $z_4$ at $-0.5$. Since $w_4 = (1, 1)$,
this split corresponds to the linear inequality $x + y \leq
-0.5$. Now, half the good states are represented in the left child of
the root (corresponding to $z_4 \leq -0.5$). The right child contains
all the bad states and the other half of the good states. So the
algorithm leaves the left child as is and tries to find the best split
for the right child. Again, we see the same pattern with $z_4$, and so
the algorithm picks the split $z_4 < 1$. Now, all bad states fall
into the left child, and all good states fall into the right
child, and we are done. The computed DT is shown in Fig.~\ref{fig:example-dtree}.

Finally, we need to convert the DT back into a formula. To do this, we can
follow all paths from the root that lead to good leaves, and take the
conjunction of all inequalities on the path, and finally take the disjunction of
all such paths. In our example, we get the candidate invariant:
\[ (x + y \leq -0.5) \vee ( x + y > - 0.5 \wedge x + y > 0.5) \enspace.\]
This can be simplified to $(x + y \neq 0)$.

\smartparagraph{Verifying the Candidate Invariant.}
Our program is then annotated with this invariant and passed to
a theorem prover to verify that the invariant is indeed sufficient to
prove the program correct.

%%% Local Variables:
%%% mode: latex
%%% TeX-master: "main"
%%% End:

%% file: preliminaries.tex
\section{Preliminaries}
\label{sec:preliminaries}

\smartparagraph{Problem Statement.}
We think of programs as transition systems $P = (S, R, \init, \safe)$,
where $S$ is a set of states, $R \subseteq S \times S$ is a
%\todo{effectively computable?}
transition relation on the states, $\init \subseteq S$ is a set of
initial states, and $\safe \subseteq S$ is the set of safe states that we want
our program to remain within. %\todo{should we define unsafe states instead?}

For any set of states $X$, the set $\post(X)$ represents all the successor
states with respect to $R$. More formally,
\[\post(X) = \pset{x'}{\exists x \in X.\, R(x, x')} \enspace. \]
Then, we can define the set of reachable states of the program (the \emph{good}
states) to be the least fixed point,
\[\good = \lfp( \l X.\, \init \cup \post(X)) \enspace. \]
Similarly, we can define
\begin{align*}
& \pre(X) = \pset{x}{\exists x' \in X.\, R(x, x')} \quad \text{and}
\quad \bad = \lfp( \l X.\, \error \cup \pre(X)),
\end{align*}
where $\error$ is the complement of $\safe$.
Then we see that the program is correct, with respect to the safety property
given by $\safe$, if $\good \cap \bad = \emptyset$. Thus, our task is to separate the
good states from the bad states.
%However, in general, calculating the set of
%reachable states of a program is undecidable.

One method to show that these sets are disjoint is to show the
existence of a \emph{safe inductive invariant}. A safe inductive
invariant is a set $\inv$ that satisfies the following three properties:
\begin{itemize}
\item $\init \subseteq \inv$,
\item $\post(\inv) \subseteq \inv$,
\item $\inv \subseteq \safe$.
\end{itemize}

It is easy to see that these conditions imply, in particular, that $\inv$
separates the good and bad states: $\good \subseteq \inv$ and $\inv \cap \bad =
\es$.

\smartparagraph{Invariant Generation as Binary Classification.}
The set up for most machine learning problems is as follows. We have an input
space $\Xx$ and an output space $\Yy$, and one is given a set of samples $X
\subseteq \Xx$ that are labeled by some unknown function $f: \Xx \to \Yy$. We
fix a hypothesis set $\Hh \subseteq \Yy^{\Xx}$, and the aim is to find the
hypothesis $h \in \Hh$ that most closely approximates $f$. The samples are often
given in terms of \emph{feature vectors}, and thus a sample $\vec{x}$ can be
thought of as a point in some $d$-dimensional space.

Binary classification is a common instance of this problem, where the
labels are restricted to a binary set $\Yy=\set{0,
  1}$. Following~\cite{SGHAN13}, we view the problem of computing a
safe inductive invariant for a program $P = (S, R, \init, \safe)$ as a
binary classification problem by defining the input space as the set
of all program states $\Xx = S$. Sample states $\vec{x} \in \bad$ are
labeled by $0$ and $\vec{x} \in \good$ are labeled by $1$.
Thus, the unknown function $f$ is the
characteristic function of the set of good states.  The hypothesis to
be learned is a safe inductive invariant. Note that if sampling the
program shows that some state $\vec{x}$ is both $\good$ and $\bad$,
there exists no safe inductive invariant, and the program is shown to
be unsafe. Hence, we assume that the set of sample states can be
partitioned into good and bad states.

\smartparagraph{Decision Trees.}
Instead of considering a hypothesis space that can represent arbitrary
invariants, we restrict it to a specific abstract domain, namely those
invariants that can be represented using decision trees.  A Decision
Tree (DT)~\cite{DBLP:books/daglib/0087929} is a binary tree that represents a Boolean
function. Each inner node $v$ of $T$ is labeled by a decision of the
form $x_i \leq t$, where $x_i$ is one of the input features and $t$ is
a real valued threshold. We denote this inequality by
$v.\mathsf{cond}$. We denote the left and right children of an inner
node by $v.\mathsf{left}$ and $v.\mathsf{right}$ respectively. Each leaf
$v$ is labeled by an output $v.\mathsf{label}$. To evaluate an input,
we trace a path down from the root node $T.\mathsf{root}$ of the tree,
going left at each inner node if the decision is true, and right
otherwise. The output of the tree on this input is the label of the leaf reached
by this process. The hypothesis set corresponding to all DTs is thus
arbitrary Boolean combinations of linear inequalities of the form $x_i
\leq t$ (axis-aligned hyperplanes).

As one can easily see, many DTs can represent the same underlying
function. However, the task of finding the smallest (in terms of number of
nodes) DT for a particular function can be shown to be NP-complete
\cite{DBLP:journals/ipl/HyafilR76}. Standard algorithms to learn DTs work by greedily
selecting at each node the co-ordinate and threshold that separates the
remaining training data best \cite{Quinlan93, DBLP:books/wa/BreimanFOS84}. This procedure is followed
recursively until all leaves have samples labeled by a single class.

The criterion for separation is normally a measure such as conditional
entropy. Entropy is a commonly used measure of uncertainty in a system. It is a
function that is low when the system is homogeneous (in this case, think of when
all samples reaching a node have the same label), and high
otherwise. Conditional entropy, analogously, measures how homogeneous the samples
are after choosing a particular co-ordinate and threshold. More formally,
at each node, we look at the samples that reach that node, and
define the conditional entropy of splitting feature $x_i$ at threshold $t$ as
\begin{align*}
  H(y | x_i : t) = {} & p(x_i < t) H(y | x_i < t) + p(x_i \geq t) H(y | x_i \geq t),\\
\text{where} \quad
H(y | x_i < t) = {} & - \sum_{a \in \Yy} p(y = a | x_i < t) \log p(y =
a | X_i < t),
\end{align*}
and $p(A)$ is the empirical probability,
i.e., the fraction of sample points reaching this node that satisfy the
condition $A$. $H(y | x_i \geq t)$ is defined similarly to $H(y | x_i < t)$.
Note that if a particular split perfectly separates good and bad samples, then
the conditional entropy is 0. The greedy heuristic is to pick the feature and
split that minimize the conditional entropy. There are other measures as well,
such as the Gini index, which is used by the DT learner we used in our
experiments~\cite{scikit-learn}.

%%% Local Variables:
%%% mode: latex
%%% TeX-master: "main"
%%% End:

%% file: algorithm.tex
\section{Algorithm}
\label{sec:algorithm}

We now present our DT-learning algorithm. We assume that we have black
box procedures for sampling points from the given program, and for
getting a set of slopes from a chosen abstract domain. To this end,
let \texttt{Sampler} be a procedure that takes a program and returns
an $n \times d$ matrix $X$ of $n$ sample points, and an
$n$-dimensional vector $\vec{y}$ corresponding to the label of each
sample point (1 for good states, 0 otherwise). Similarly, let
\texttt{Slopes} be a procedure that takes a program and returns an $m
\times d$ matrix $H$ of $m$ hyperplane slopes. We describe the actual
procedures used in our experiments in Section~\ref{sec:implementation}.

Our final algorithm is surprisingly simple, and is given in
Algorithm~\ref{fig:algorithm}. We get the sample points and the slopes
from the helper functions mentioned above, and then transform the
sample points according to the slopes given. We run a standard DT
learning algorithm on the transformed sample to obtain a tree that
classifies all samples correctly. The tree is then transformed into a
formula that is a candidate invariant, by a simple procedure
\texttt{DTtoForm}. Finally, the program is annotated with the
candidate invariant and verified. This final step is realized by
another black box procedure \texttt{IsInvariant}, which checks that
the invariant satisfies the three conditions necessary to be a safe
inductive invariant. For example, this can be done by encoding $\init,
\inv, \post $ and $\safe$ as SMT formulas and feeding the three
conditions into an SMT solver.

\newcommand{\fail}{\mathsf{fail}}

\begin{algorithm}[t]
    \LinesNumberedHidden
    \SetKwFunction{algo}{DTInv}
    \SetKwFunction{tf}{DTtoForm}
    \SetKwFunction{sampler}{Sampler}
    \SetKwFunction{slopes}{Slopes}
    \SetKwFunction{learndt}{LearnDT}
    \SetKwFunction{check}{IsInvariant}
    \SetKwData{v}{v}
    \SetKwData{lbl}{label}
    \SetKwData{cond}{cond}
    \SetKwData{left}{left}
    \SetKwData{right}{right}
    \SetKwData{root}{root}
    \SetKwProg{myfun}{Function}{}{end}
    \SetKwProg{fun}{def}{${} =$}{}
    \SetArgSty{}
    \DontPrintSemicolon
    \SetKw{Val}{val}
    \SetKwIF{If}{ElseIf}{Else}{if}{then}{else if}{else}{}

    \fun{\algo($P$ : program): safe inductive invariant for $P$ or
      $\fail$ }{
      \Val $X, \vec{y} = \sampler{P}$\;
      \Val $H = \slopes{P}$\;
      \Val $Z = X \cdot H^T$\;
      \Val $T =$ \learndt{$Z$, $\vec{y}$}\;
      \Val $\f = \tf{T.\root}$\;
      \lIf{\check{$P$, $\f$}}{$\f$} \lElse{$\fail$}

    }
    \fun{\tf(\v: node of a decision tree): formula represented by subtree rooted
      at \v}{
      \lIf{\v is a leaf}{\v.\lbl}
      \lElse{$(\v.\cond \land \tf{\v.\left} \lor \neg \v.\cond \land
          \tf{\v.\right})$ }
    }

    \caption{$\mathtt{DTInv}$: Invariant generation algorithm using DT learning}
  \label{fig:algorithm}
\end{algorithm}

To convert the DT into a formula, we note that the set of states that reach a
particular leaf is given by the conjunction of all predicates on the path from
the root to that leaf. Thus, the set of all states classified as good by the DT
is the disjunction of the sets of states that reach all the good leaves. A
simple conversion is then to take the disjunction over all paths to good leaves
of the conjunction of all predicates on such paths. The procedure
\texttt{DTtoForm} computes this formula recursively by traversing the
learned DT.
% Let \texttt{toFormula(v)}
% represent the formula corresponding to the subset of states that reach $V$ that
% are good. For leaves, this is simply \texttt{true} or \texttt{false} depending
% on the label of the leaf. Inductively, for a non-leaf node $v$, if the predicate
% on $v$ is $\f$, then states that satisfy $\f$ reach its left child, and states
% that do not reach its right child. Thus, by the inductive hypothesis, the
% subsets of states reaching $v$ that will be classified as good are $(\f \wedge
% $\texttt{toFormula($v_l$)}$)\vee (\neg \f \wedge$\texttt{toFormula($v_r$)}$)$,
% where $v_l, v_r$ are the left and right children of $v$ respectively. This
% procedure is given in Algorithm \ref{fig:toformula}.
Since the tree was learnt on the transformed sample data,
the predicate at each node of the tree will be of the form $z_i
\leq c$ where $z_i$ is one of the columns of $Z$ and $c$ is a
constant. Since $Z = X \cdot H^T$, we know that $z_i(\vec{x}) =
\vec{x} \cdot \vec{w_i}$ for a sample $\vec{x}$. Thus, the predicate
is equivalent to $\vec{x} \cdot \vec{w_i} \leq c$, which is a linear
inequality in the program variables. Combining this with the
conversion procedure above, we see that our algorithm outputs an
invariant which is a Boolean combination of linear inequalities over
the numerical program variables.

\smartparagraph{Soundness.}  From the above discussion, we see that
the formula $\f$ returned by the procedure \texttt{DTtoForm} is of the
required format. Moreover, we assume that the procedure
\texttt{IsInvariant} is correct. Thus, we can say that our invariant
generation procedure \texttt{DTInv} is sound: if it terminates
successfully, it returns a safe inductive invariant.

\smartparagraph{Probabilistic Completeness.}
It is harder to prove that such invariant generation algorithms are complete. We
see that the performance of our algorithm depends heavily on the sample set, if
the sample is inadequate, it is impossible for the DT learner to learn the
underlying invariant. One could augment our algorithm with a refinement loop,
which would make the role of the sampler less pronounced, for if the invariant
is incorrect, the theorem prover will return a counterexample that could
potentially be added to the sample set and we can re-run learning. However, we
find in practice that we do not need a refinement loop if our sample set is
large enough.

We can justify this observation using Valiant's \emph{PAC}
(probably approximately correct) model \cite{Valiant84}. In this
model, one can prove that an algorithm that classifies a large enough
sample of data correctly has small error on all data, with high
probability. It must be noted that one key assumption of this model is
that the sample data is drawn from the same distribution as the
underlying data, an assumption that is hard to justify in most of its
applications, including this one. In practice however, PAC learning
algorithms are empirically successful on a variety of applications
where the assumption on distribution is not clearly true.
Formally, we can give a generalization guarantee for our algorithm using this
result of Blumer et al. \cite{DBLP:journals/jacm/BlumerEHW89}:
\begin{theopargself}
\begin{theorem}
\label{thm:blumer}
A learning algorithm for a hypothesis class $\Hh$ that outputs a
hypothesis $h \in \Hh$ will have true error at most $\e$ with
probability at least $1 - \d$ if $h$ is consistent with a sample of
size $\max (\frac{4}{\e}\log\frac{2}{\d},
\frac{8VC(\Hh)}{\e}\log\frac{13}{\e})$.
\end{theorem}
\end{theopargself}
In the above theorem, a hypothesis is said to be \emph{consistent}
with a sample if it classifies all points in the sample correctly. The
quantity $VC(\Hh)$ is a property of the hypothesis class called the
Vapnik-Chervonenkis (VC) dimension, and is a measure of the
expressiveness of the hypothesis class. As one might imagine, more
complex classes lead to a looser bound on the error, as they are more
likely to over-fit the sample and less likely to generalize well.

Thus, it suffices for us to bound the VC dimension of our hypothesis
class, which is all finite Boolean combinations of hyperplanes in $d$
dimensions. The VC dimension of a class $\Hh$ is defined as the
cardinality of the largest set of points that $\Hh$ can shatter. A set
of points is said to be \emph{shattered} by a hypothesis class $\Hh$
if for every possible labeling of the points, there exists a
hypothesis in $\Hh$ that is consistent with it. Unfortunately, DTs in
all their generality can shatter points of arbitrarily high
cardinality. Given any set of $m$ points, we can construct a DT with
$m$ leaves such that each point ends up at a different leaf, and now
we can label the leaf to match the labeling given.

Since in practice we will not be learning arbitrarily large trees, we can
restrict our algorithm a-priori to stop growing the tree when it reaches $K$
nodes, for some fixed $K$ independent of the sample. Now one can use a basic, well-known lemma
from \cite{Mohri:homework} combined with Sauer's Lemma \cite{Sauer} to get that
the VC dimension of bounded decision trees is $O(Kd\log K)$. Combining this with
Theorem \ref{thm:blumer}, we get the following polynomial bound for probabilistic
completeness:

\begin{theorem}
  Under the assumptions of the PAC model, the algorithm \texttt{DTInv} returns
  an invariant that has true error at most $\e$ with probability at least $1 -
  \d$, given that its sample size is $O(\frac{1}{\e}Kd\log K \log \frac{1}{\d})$ .
\end{theorem}

%Note that the sample size required is polynomial in $K$ and $d$.

\smartparagraph{Complexity.}
The running time of our algorithm depends on many factors such as
the running time of the sampling (which in turn depends on the benchmark being
considered), and so is hard to measure precisely. However, the running time of
the learning routine for DTs is $O(m n \log(n))$, where $m$ is the number of
hyperplane slopes in $H$ and $n$ is the number of sample points \cite{DBLP:books/wa/BreimanFOS84, scikit-learn}. The learning algorithm therefore scales well to
large sets of sample data.
\iffalse
This compares favorably with the set-cover algorithm of
\cite{SGHAN13}, where the running time of the learning routine is $O(m n^3)$.
This is because the greedy algorithm for set cover takes time $O( h n^2)$ where
$h$ is the number of hyperplanes considered, and since they take the set of
candidate slopes and consider one hyperplane for every point with every slope,
they have $h = m n$. Thus we see that our algorithm is able to handle larger
sample sets with ease.
\fi

\smartparagraph{Nonlinear invariants.}
An important property of our algorithm is that it generalizes elegantly to
nonlinear invariants as well. For example, if a particular program requires
invariants that reason about $(x \mod 2)$ for some variable $x$, then we can learn
such invariants as follows: given the sampled states $X$, we add to it a new
column that corresponds to a variable $x'$, such that $x' = x \mod 2$. We then
run the rest of our algorithm as before, but in the final invariant, replace all
occurrences of $x'$ by $(x \mod 2)$. As is easy to see, this procedure correctly
learns the required nonlinear invariant. We have added this feature to our
implementation, and show that it works on benchmarks requiring these nonlinear
features (see Section \ref{sec:evaluation}).

%%% Local Variables:
%%% mode: latex
%%% TeX-master: "main"
%%% End:

%% file: results.tex
\section{Implementation and Evaluation}

\subsection{Implementation}
\label{sec:implementation}

We implemented our algorithm in Python, using the
\texttt{scikit-learn} library's decision tree
classifier~\cite{scikit-learn} as the DT learner
\texttt{LearnDT}. This implementation uses the CART algorithm from
\cite{DBLP:books/wa/BreimanFOS84} which learns in a greedy manner as described in
Section~\ref{sec:preliminaries}, and uses the Gini index.

We implemented a simple and naive \texttt{Sampler}: we considered all
states that satisfied the precondition where the value of every variable was in
the interval $[-L, L]$. For these states, we ran the program with a bound $I$ on the
number of iterations of loops, and collected all states reached as good
states. To find bad states, we looked at all states that were a margin $M$ away
from every good state, ran the program from this state (again with at most $I$
iterations), and if the program failed an assertion, we collected all the states
on this path as bad states. The bounds $L, I, M$ were initialized to low values
and increased until we had sampled enough states to prove our program correct.

For the \texttt{Slopes} function, we found that for most of the
programs in our benchmarks it sufficed to consider slopes in the
\emph{octagonal} abstract domain. This consists of all vectors in
$\{-1, 0, 1\}^d$ with at most two non zero elements. In a few cases,
we needed additional slopes (see Table \ref{tbl:results}). For the
programs \texttt{hola15} and \texttt{hola34}, we used a class of
slopes of vectors in $C^d$ where $C$ is a small set of constants that
appear in the program, and their negations.

We also developed methods to learn the class of slopes needed by a
program by looking at the states sampled. In \cite{SGHAN13}, the
authors suggest working in the \emph{null space} of the good states,
viewed as a matrix. This is because if the good states lie in some
lower dimensional space, this would automatically suggest equality
relationships among them that can be used in the invariant. It also
reduces the running time of the learning algorithm. Inspired by this,
we propose using \emph{Principal Component Analysis}~\cite{PCA:book} on the
good states to generate slopes. PCA is a method to find the basis of a
set of points so as to maximize the variance of the points with
respect to the basis vectors. For example, if all the good points lie
along the line $2 x + 3 y = 4$, then the first PCA vector will be $(2,
3)$, and intuition suggests that the invariant will use inequalities
of the form $2 x + 3 y < c$. %\todo{Confirm this. Say anything more?}

Finally, for the \texttt{IsInvariant} routine, we used the program verifier
Boogie~\cite{Boogie}, which allowed us to annotate our programs with the
invariants we verified. Boogie uses the SMT solver Z3~\cite{z3} as a back-end.

\subsection{Evaluation}
\label{sec:evaluation}

\newcommand{\tool}{\texttt}

We compared our algorithm with a variety of other invariant inference
tools and static analyzers. We mainly focused on ML-based
algorithms, but also considered tools based on interpolation and
abstract interpretation. Specifically, we considered:
\begin{itemize}
\item \tool{ICE}~\cite{DBLP:conf/cav/0001LMN14}: an ML algorithm
  based on ICE-learning that uses an SMT solver to learn numerical invariants.
\item \tool{MCMC}~\cite{DBLP:conf/cav/0001A14}: an ML algorithm based
  on Markov Chain Monte Carlo methods. There are two versions of this
  algorithm, one that uses templates such as octagons for the invariant,
  and one with all constants in the slopes picked from a fixed bag of
  constants. The two algorithms have very similar characteristics. We ran both
  versions 5 times each (as they are randomized) and report the better
  average result of the two algorithms for each benchmark.
\item \tool{SC}~\cite{SGHAN13}: an ML algorithm based on set cover. We only had
  access to the learning algorithm proposed in~\cite{SGHAN13} and not the
  sampling procedure. To obtain a meaningful comparison, we combined it with the
  same sampler that we used in the implementation of our algorithm \tool{DTInv}.
\item \tool{CPAchecker}~\cite{DBLP:conf/cav/BeyerK11}: a
  configurable software
  model checker. We chose the default analysis based on predicate abstraction and interpolation.
\item \tool{UFO}~\cite{DBLP:conf/tacas/AlbarghouthiGLCC13}: a
  software model checker that combines abstract interpretation and interpolation
  (denoted \tool{CPA}).
\item \tool{InvGen}~\cite{DBLP:conf/cav/GuptaR09}: an inference tool
  for linear invariants that combines abstract interpretation,
  constraint solving, and testing.
\end{itemize}
For our comparison, we chose a combination of 22 challenging
benchmarks from various sources. In particular, we considered a subset
of the benchmarks from~\cite{SGHAN13, DBLP:conf/cav/0001LMN14,
  DBLP:conf/cav/GuptaR09, DBLP:conf/oopsla/DilligDLM13}. We chose the
benchmarks at random among those that were hard for at least one other
tool to solve. Due to this bias in the selection, our experimental
results do not reflect the average performance of the tools that we
compare against. Instead, the comparison should be considered as an
indication that our approach provides a valuable complementary
technique to existing algorithms.

\begin{table}[!t]
\centering
\input{table.tex}
\caption{Results of comparison. The table is divided according to what kinds of
  invariants the benchmark needed. The
  ``Type'' column denotes the Boolean structure of the invariant - conjunctive,
  disjunctive and arbitrary Boolean combination are denoted as conj., disj. and ABC
  respectively. The column $|\f|$ contains the number of predicates in the
  invariant found by \texttt{DTInv}.\\
  The columns 'Samp' and 'DT' show the running time in seconds of our
  sampling and DT learning procedures respectively. We show \tool{SC} next, as
  we only compare learning times with \tool{SC}. Then follows the total time of our
  tool (\tool{DTInv}), followed by those for other tools.
  Each entry of the tool columns shows the
  running time in seconds if a safe invariant was found, or otherwise
  one of the following entries. 'NA': program contains arithmetic
  operations that are not supported by the tool; 'F': analysis terminated without
  finding a safe invariant; 'TO': timeout; 'MO': out of
  memory. The times for \tool{MCMC} have an asterisk if at least one of the
  repetitions timed out. In this case the number shown is the average of the
  other runs.
  \label{tbl:results}}
\end{table}

We ran our experiments on a machine with a quad-core 3.40GHz CPU and
16GB RAM, running Ubuntu GNU/Linux. For the analysis of each benchmark, we
used a memory limit of 8GB and a timeout of 5 minutes. The results of
the experiment are summarized in Table~\ref{tbl:results}.
Here are the observations from our experiments (we provide more
in-depth explanations for these observations in the next section,
where we discuss related work in more detail):
\begin{itemize}
\item Our algorithm \tool{DTInv} seems to learn complex Boolean invariants as
  easily as simple conjunctions.
\item \tool{ICE} seems to struggle on programs that needed large invariants. For
  some of these (gopan, popl07), this is because the constraint solver runs out
  of time/memory, as the constants used in the invariants are also large. For
  \texttt{hola19} and \texttt{prog4}, ICE stops because the tool has an inbuilt
  limit for the complexity of Boolean templates.
\item However, \tool{ICE} solves \texttt{sum1} and \texttt{trex3} quickly, even
  though they need many predicates, because the constant terms are small, so
  this space is searched first by \tool{ICE}.
\item Similarly, we notice that \tool{MCMC} has difficulty finding large
  invariants, again because the search space is huge.
\item \tool{SC}'s learning algorithm is consistently slower than \tool{DTInv},
  due to its higher running time complexity. It also runs out of memory for large
  sample sizes.
\item \tool{DTInv} is able to easily handle benchmarks that \tool{CPA},
  \tool{UFO} and \tool{InvGen} struggle on. This is mainly because they are
  specialized for reasoning about linear invariants, and have issues dealing
  with invariants that have complicated Boolean structure.
\end{itemize}

We also learned some of the weak points in our current approach:
\begin{itemize}
\item \tool{DTInv} is slow in processing \texttt{fig1} and \texttt{prog4}, both
  of which are handled by at least one other tool without much effort. However,
  we note that most of this time is spent in the sampling routine, which is
  currently a naive implementation. We therefore believe that DT learning could
  benefit from a combination with a static analysis that provides approximations
  of the good and bad states to guide the sampler.
\item We also note that the method of ``constant slopes'' which we
  used to handle the non octagonal benchmarks (\texttt{hola15}, \texttt{hola34}) is
  ad-hoc, and might not work well for larger benchmarks.
\end{itemize}

\smartparagraph{Beyond octagons.} As mentioned in Section~\ref{sec:algorithm},
we implemented a feature to learn certain nonlinear invariants. We were able to
verify some benchmarks that needed reasoning about the modulus of certain
variables, as shown in Table~\ref{tbl:results}. Finally, we show one example
where we were able to infer a nonlinear invariant (specifically $s = i^2 \wedge
i \leq n$ for \texttt{square}).

We believe our experiments show that Decision Trees are a natural representation
for invariants, and that the greedy learning heuristics guide the algorithm to
discover simple invariants of complex structures without additional overhead.

%%% Local Variables:
%%% mode: latex
%%% TeX-master: "main"
%%% End:

%% file: table.tex
\setlength\extrarowheight{1pt}
\begin{tabular}{|l|c|c|c||r|r|r||>{\bfseries}r|r|r|r|r|r|}
 \hline
\textbf{Name} & \textbf{Vars} & \textbf{Type} & $\boldsymbol{|\f|}$ &
\textbf{Samp} & \textbf{DT} & \textbf{SC} & \textbf{DTInv} &
\textbf{ICE} & \textbf{MCMC} & \textbf{CPA} & \textbf{UFO} & \textbf{InvGen}\\
\hline
\multicolumn{11}{l}{Octagonal:}\\
\hline
ex23\cite{DBLP:conf/cav/0001LMN14} & 4 & conj. & 3 & 0.10 & 0.01 & 4.73 & 0.11 & 8.82 & 0.01 & 19.77 & 1.50 & 0.02\\
fig6\cite{DBLP:conf/cav/0001LMN14} & 2 & conj. & 2 & 0.00 & 0.00 & 2.61 & 0.00 & 0.30 & 0.00 & 1.68 & 0.13 & 0.01\\
fig9\cite{DBLP:conf/cav/0001LMN14} & 2 & conj. & 2 & 0.00 & 0.01 & 2.31 & 0.01 & 0.33 & 0.00 & 1.73 & 0.13 & 0.01\\
hola10\cite{DBLP:conf/oopsla/DilligDLM13} & 4 & conj. & 8 & 0.03 & 0.00 & 2.45 & 0.03 & 49.21 & TO & 2.03 & F & F\\
nested2\cite{SGHAN13} & 4 & conj. & 3 & 0.03 & 0.00 & 2.39 & 0.03 & 62.02 & 0.09 & 1.86 & 0.12 & 0.03\\
nested5\cite{DBLP:conf/cav/GuptaR09} & 4 & conj. & 4 & 2.48 & 0.02 & MO & 2.50 & 60.95 & 31.28* & 2.08 & 0.35 & 0.03\\
fig1\cite{DBLP:conf/cav/0001LMN14} & 2 & disj. & 3 & 14.61 & 0.01 & F & 14.62 & 0.38 & 5.13 & 1.75 & 1.64 & F\\
test1\cite{SGHAN13} & 4 & disj. & 2 & 0.90 & 0.01 & 7.86 & 0.91 & 0.39 & TO & 1.71 & F & 0.04\\
cegar2\cite{DBLP:conf/cav/0001LMN14} & 3 & ABC & 5 & 0.03 & 0.01 & 2.64 & 0.04 & 4.86 & 17.30 & 1.97 & 0.18 & F\\
gopan\cite{SGHAN13} & 2 & ABC & 8 & 0.03 & 0.00 & 2.54 & 0.03 & F & TO & 63.85 & 58.29 & F\\
hola18\cite{DBLP:conf/oopsla/DilligDLM13} & 3 & ABC & 6 & 1.60 & 0.04 & MO & 1.64 & TO & 21.93* & TO & 8.38 & F\\
hola19\cite{DBLP:conf/oopsla/DilligDLM13} & 4 & ABC & 7 & 0.19 & 0.01 & 3.47 & 0.20 & F & TO & F & F & F\\
popl07\cite{SGHAN13} & 2 & ABC & 7 & 0.03 & 0.00 & 2.72 & 0.03 & F & TO & 110.81 & 15.20 & F\\
prog4\cite{SGHAN13} & 3 & ABC & 8 & 2.32 & 0.02 & MO & 2.34 & F & 0.13 & F & F & F\\
sum1\cite{DBLP:conf/cav/0001LMN14} & 3 & ABC & 6 & 0.01 & 0.01 & 2.61 & 0.02 & 1.32 & 29.04* & F & 0.17 & F\\
trex3\cite{DBLP:conf/cav/0001LMN14} & 8 & ABC & 9 & 8.44 & 0.06 & MO & 8.50 & 4.51 & NA & F & 0.18 & F\\
\hline
\multicolumn{11}{l}{Non octagonal:}\\
\hline
hola15\cite{DBLP:conf/oopsla/DilligDLM13} & 3 & conj. & 2& 0.52 & 0.02 & 0.01 & 0.54 & 0.53 & 0.04 & MO & 0.13 & 0.02\\
\hline
\multicolumn{11}{l}{Modulus:}\\
\hline
hola02\cite{DBLP:conf/oopsla/DilligDLM13} & 4 & conj. & 4 & 0.03 & 0.03 & F &
0.06 & F & NA & F & F & F\\
hola06\cite{DBLP:conf/oopsla/DilligDLM13} & 4 & conj. & 3 & 1.79 & 0.03 & MO &
1.82 & F & NA & F & F & F\\
hola22\cite{DBLP:conf/oopsla/DilligDLM13} & 4 & conj. & 5 & 0.04 & 0.00 & F &
0.04 & F & NA & F & F & F\\
\hline
\multicolumn{11}{l}{Non octagonal modulus:}\\
\hline
hola34\cite{DBLP:conf/oopsla/DilligDLM13} & 4 & ABC & 6 & 1.14 & 0.03 & 3.10 & 1.17 & F & NA & F & F & F\\
\hline
\multicolumn{11}{l}{Quadratic:}\\
\hline
square & 3 & conj. & 3 & 0.27 & 0.24 & MO & 0.51 & F & NA & F & F & F\\
\hline
\end{tabular}

%%% Local Variables:
%%% mode: latex
%%% TeX-master: "main"
%%% End:

%% file: related.tex
\section{Related Work and Conclusions}

%\todo{Perhaps mention Daikon and related tools}

Our experimental evaluation compared against other algorithms for
invariant generation. We discuss these algorithms in more
detail. Sharma et.\ al.~\cite{SGHAN13} used the greedy set cover
algorithm \tool{SC} to learn invariants in the form of arbitrary
Boolean combinations of linear inequalities. Our algorithm based on
decision trees is simpler than the set cover algorithm, works better
on our benchmarks (which includes most of the benchmarks
from~\cite{SGHAN13}), and scales much better to large sample sets of
test data. The improved scalability is due to the better complexity of
DT learners. The running time of our learning algorithm is
$\mathcal{O}(m n \log(n))$ where $m$ is the number of
features/hyperplane slopes that we consider, and $n$ the number of
sample points. On the other hand, the set cover algorithm has a
running time of $\mathcal{O}(m n^3)$. This is because the greedy
algorithm for set cover takes time $\mathcal{O}(h n^2)$ where $h$ is
the number of hyperplanes, and~\cite{SGHAN13} considers one hyperplane
for every candidate slope and sample point, yielding $h = m n$.

When invariant generation is viewed as binary classification, then
there is a problem in the refinement loop: if the learned invariant is
not inductive, it is unclear whether the counterexample model produced
by the theorem prover should be considered a ``bad'' or a ``good''
state. The ICE-learning framework~\cite{DBLP:conf/cav/0001LMN14}
solves this problem by formulating invariant generation as a more
general classification problem that also accounts for implication
constraints between points. We note that our algorithm does not fit
within this framework, as we do not have a refinement loop that can
handle counterexamples in the form of implications. However, we found
in our experiments that we did not need any refinement loop as our
algorithm was able to infer correct invariants directly after sampling
enough data. Nevertheless, considering an ICE version of DT learning is
interesting as sampling without a refinement loop becomes difficult
for more complex programs.

The paper~\cite{DBLP:conf/cav/0001LMN14} also proposes a concrete
algorithm for inferring linear invariants that fits into the
ICE-learning framework (referred to as \tool{ICE} in our
evaluation). If we compare the complexity of learning given a fixed
sample, our algorithm performs better than
\cite{DBLP:conf/cav/0001LMN14} both in terms of running time and
expressiveness of the invariant. The ICE algorithm of
\cite{DBLP:conf/cav/0001LMN14} iterates through templates for the
invariant. This iteration is done by dovetailing between more complex
Boolean structures and increasing the range of the thresholds
used. For a fixed template, it formulates the problem of this template
being consistent with all given samples as a constraint in
quantifier-free linear integer arithmetic. Satisfiability of this
constraint is then checked using an SMT solver. We note that the size
of the generated constraint is linear in the sample size, and that
solving such constraints is NP-complete. In comparison, our learning
is sub-quadratic time in the sample size. Also, we do not need to fix
templates for the Boolean structure of the invariant or bound the
thresholds a priori. Instead, the DT learner automatically infers
those parameters from the sample data.

Another ICE-learning algorithm based on randomized search was proposed
in~\cite{DBLP:conf/cav/0001A14} (the algorithm \tool{MCMC} in our
evaluation). This algorithm searches over a fixed space of invariants
$S$ that is chosen in advance either by bounding the Boolean structure
and coefficients of inequalities, or by picking some finite
sub-lattice of an abstract domain. Given a sample, it randomly
searches using a combination of random walks and hill climbing until
it finds a candidate invariant that satisfies all the samples. There
is no obvious bound on the time of this search other than the trivial
bound of $|S|$. Again, we have the advantage that we do not have to
provide templates of the Boolean structure and the thresholds of the
hyperplanes. These parameters have to be fixed for the algorithm
in~\cite{DBLP:conf/cav/0001A14}.  Furthermore, the greedy nature of DT
learning is a heuristic to try simpler invariants before more complex
ones, and hence the invariants we find for these benchmarks are often
much simpler than those found by MCMC.

Decision trees have been previously used for inferring likely
preconditions of
procedures~\cite{DBLP:conf/issta/SankaranarayananCIG08}. Although this
problem is related to invariant generation, there are considerable
technical differences to our algorithm. In particular, the algorithm
proposed in~\cite{DBLP:conf/issta/SankaranarayananCIG08} only learns
formulas that fall into a finite abstract domain (Boolean
combinations of a given finite set of predicates), whereas we use
decision trees to learn more general formulas in an
infinite abstract domain (e.g., unions of octagons).

We believe that the main value of our algorithm is its ability to
infer invariants with a complex Boolean structure efficiently from
test data. Other techniques for inferring such invariants include
predicate
abstraction~\cite{GrafSaidi97ConstructionAbstractStateGraphsPVS} as
well as abstract interpretation techniques such as disjunctive
completion~\cite{CousotCousot79SystematicDesignProgramAnalysis}.
However, for efficiency reasons, many static analyses are restricted
to inferring conjunctive invariants in
practice~\cite{DBLP:conf/tacas/BallPR01,
  DBLP:conf/esop/CousotCFMMMR05}. There exist techniques for
recovering loss of precision due to imprecise joins using
counterexample-guided
refinement~\cite{PodelskiWies10CounterexampleGuidedFocus,
  DBLP:conf/sas/AlbarghouthiGC12, tarefinement}. In the future, we
will explore whether DL learning can be used to complement such
refinement techniques for static analyses.

%Bias-variance trade-offs~\cite{DBLP:conf/popl/0001NA14}.

%%% Local Variables:
%%% mode: latex
%%% TeX-master: "main"
%%% End: